\DocumentMetadata{
	lang		= eng-US, 	
	pdfstandard	= A-3u,		
	pdfversion	= 1.7, 		
}

\documentclass[colorlinks,upint,subscriptcorrection,varvw,hyphenate,balance,greek,russian,vietnamese,german]{asmeconf} 
\usepackage{algorithm}
\usepackage{algpseudocode}

\usepackage[inkscapelatex=false]{svg}
\usepackage{float}

\hypersetup{%
	pdfauthor={Uzair Aziz Muhammad},									  
	pdftitle={Multi-Axis Additive Manufacturing for Customized Automotive Components},                  
	pdfkeywords={Additive manufacturing, digital light processing, smart manufacturing, vat photopolymerization},
	pdfsubject = {asmeconf},			  
}

\allowdisplaybreaks 

\begin{document}


\ConfName{Proceedings of the ASME 2026\linebreak Global Transportation and Power Systems Summit}
\ConfAcronym{DRIVN2026}
\ConfDate{September 14--16, 2026} 
\ConfCity{Detroit, MI}
\PaperNo{DRIVN2026-191681}

%

\title{Multi-Axis Additive Manufacturing for Customized Automotive Components} 
 
%
%
%



\SetAuthors{%
  Uzair Aziz Muhammad\affil{1},%
  Zheng Liu\affil{2}\CorrespondingAuthor{zhengtl@umich.edu}%
}
\SetAffiliation{1}{Department of Computer and Information Science,
  University of Michigan-Dearborn, Dearborn, Michigan, USA}
\SetAffiliation{2}{Department of Industrial and Manufacturing Systems Engineering,
  University of Michigan-Dearborn, Dearborn, Michigan, USA}


\maketitle





\begin{abstract}
The reproduction of automobile components through additive manufacturing presents significant geometric challenges, as many automotive parts feature complex, organically shaped surfaces that are difficult to fabricate accurately using conventional 3D printing approaches without wasteful support structures. Multi-axis Digital Light Processing (DLP) 3D printing addresses this by orienting a robotic arm to cure resin layers at varying angles and positions, enabling the fabrication of geometries that fixed-axis systems cannot reliably reproduce. However, this flexibility introduces a key challenge: layers printed at non-orthogonal orientations exhibit non-uniform thickness across their cross-section, which traditional DLP systems cannot accommodate without subdividing the layer, increasing total layer count, print time, and the need for supporting structures. This paper introduces a variable exposure method to address this challenge. Rather than splitting a non-uniform layer into multiple uniform ones, our approach divides each layer into sublayers and modulates the UV illumination duration for each sublayer proportionally to its local thickness. This is governed by an established cure-depth equation relating exposure time to material penetration depth, allowing precise control over curing without additional hardware. The result is a meaningful reduction in total layer count for printed objects. Fewer layers directly translates to faster print times and a reduction in wasteful support structures. Our contribution is a practical and low-overhead extension to existing multi-axis DLP pipelines that improves print efficiency without sacrificing geometric accuracy, reducing the resin usage over 47\%.
\end{abstract}


\begin{nomenclature}

\entry{$M$}{Triangle mesh}
\entry{$V$}{Vertex set}
\entry{$F$}{Face set}
\entry{$N$}{Number of layers}
\entry{$k$}{Layer index}
\entry{$\phi(x,y,z)$}{Scalar field over mesh vertices}
\entry{$\phi_{min}, \phi_{max}$}{Scalar field range bounds}
\entry{$iso_k$}{Iso-value of layer $k$}
\entry{$P_k$}{Polyline contours for layer $k$}
\entry{$S_k$}{Set of edge-crossing segments for layer $k$}
\entry{$\varepsilon$}{Intersection epsilon tolerance}
\entry{$\delta_{weld}$}{Point welding tolerance}
\entry{$\alpha$}{Linear interpolation parameter for edge crossings}
\entry{$A$}{Warped-$z$ field amplitude}
\entry{$c_x, c_y$}{Warped-$z$ field cycle parameters}
\entry{$\hat{x}, \hat{y}$}{Bounding-box normalized coordinates}
\entry{$\mathbf{c}$}{Center point for radial scalar field}

\entry{$C_d$}{Cure depth}
\entry{$D_p$}{Penetration depth of the resin}
\entry{$H_e$}{Photoexposure dose}
\entry{$H_0$}{Critical photoexposure dose required to initiate polymerization}
\entry{$I_0$}{Peak irradiance of the projector}
\entry{$t_e$}{Exposure time}

\entry{$u, v$}{Normalized projector pixel coordinates}
\entry{$h(u,v)$}{Target cure depth at pixel $(u,v)$}
\entry{$I(u,v)$}{Effective irradiance at pixel $(u,v)$}
\entry{$G(u,v)$}{Grayscale command at pixel $(u,v)$}
\entry{$W, H$}{Cure mask width and height in pixels}
\entry{$N_{exp,k}$}{Number of interior (exposed) pixels in layer $k$}
\entry{$N_{sat,k}$}{Number of saturated pixels in layer $k$}
\entry{$p$}{Projector pixel pitch [mm/pixel]}
\entry{$A_k$}{Physical exposed area of layer $k$}
\entry{$\sigma_k$}{Saturation ratio for layer $k$}
\end{nomenclature}


\section{Introduction}

The automotive industry increasingly demands fabrication methods capable of producing complex, geometrically intricate components with minimal waste and reduced lead times. Traditional subtractive manufacturing processes, while mature and reliable, struggle to accommodate the organic surfaces and internal geometries common in modern automotive part design without significant material loss and tooling overhead. Additive manufacturing (AM) has emerged as a compelling alternative, enabling the direct fabrication of complex geometries layer by layer from digital models \cite{patel2017highly}. Its ability to support rapid prototyping and on-demand production makes it particularly well-suited for automotive applications, where design iteration and part customization are frequent requirements \cite{dilberoglu2017role}.

Among the various AM technologies available, Digital Light Processing (DLP) 3D printing stands out for its combination of high resolution and rapid fabrication speed \cite{liu2022acoustophoretic,li2021digital,wang20203d}. Unlike filament-based methods such as Fused Deposition Modeling (FDM), DLP cures entire cross-sectional layers of photopolymer resin simultaneously using a projected UV light source, enabling fast and precise deposition. However, conventional DLP systems rely on a fixed, planar build orientation, which limits their ability to fabricate overhanging geometries without the use of wasteful support structures \cite{jumbo2021digital}. These supports increase material consumption, extend post-processing time, and compromise surface quality upon removal, all significant drawbacks in a manufacturing context \cite{maines2021sustainable,liu2025uncertainty,piedra20213d,liu2022design}.

To address these limitations, researchers have explored multi-axis approaches in AM more broadly. Multi-axis FDM systems have demonstrated the ability to reduce or eliminate support structures by dynamically reorienting the print head or build platform during fabrication \cite{dai2018support}. Fang et al. \cite{fang2020reinforced} introduced a computational framework for multi-axis FDM that decomposes a solid model into a sequence of collision-free working surfaces, generating toolpaths that align filaments along stress directions for improved mechanical performance.

Further advances in multi-axis FDM have explored curved-layer slicing and volume decomposition strategies to improve both surface quality and structural integrity. Feng et al. \cite{feng2021curved} developed a curved-layer material extrusion framework coupling conformal surface offsetting with geodesic distance-based toolpath generation, enabling support-free deposition on thin, freeform shells. Bi et al. \cite{bi2023strength} proposed a volume decomposition method for multi-directional FDM that jointly optimizes mechanical strength and support-free manufacturability, validated through FEA simulations and multi-axis printing experiments. Geodesic field-based approaches \cite{li2021multi} have similarly been applied to enable support-free fabrication on multi-axis platforms. However, these algorithms are designed around the continuous extrusion mechanics of FDM and cannot be directly transferred to DLP printing, which relies on discrete, planar layer exposures rather than continuous toolpaths.

The integration of multi-axis platforms with DLP printing introduces unique challenges that existing FDM-based frameworks do not address. When the build platform is dynamically reoriented, sliced layers are no longer uniformly thick, regions of the same layer may require different depths of cure depending on local geometry. Conventional DLP systems apply a uniform exposure time per layer, which is insufficient to accurately cure layers of non-uniform thickness. While Liu et al. \cite{liu2026support} established a foundational framework for multi-axis DLP printing, including a modified slicing algorithm adapted from Fang et al. \cite{fang2020reinforced}, a Beer-Lambert-based cure depth model, and a collision-free path planning algorithm, the generation of spatially resolved cure maps for non-uniform layers was not fully addressed.
Grayscale modulation of UV exposure in DLP printing is not itself a new concept. Prior work has demonstrated its utility in a number of planar printing contexts: Mostafa et al.~\cite{mostafa2017tolerance} applied grayscale pixel values and exposure time variation to control dimensional tolerance in fixed-axis DLP systems, while Yu et al.~\cite{yu2023high} used localized grayscale adjustment to mitigate the optical proximity effect in planar microchannel fabrication. Luongo et al.~\cite{luongo2020microstructure} applied grayscale patterns to surface voxels to control reflectance and surface microstructure in standard planar DLP printing, and Valentincic et al.~\cite{valentinvcivc2017low} employed software masks to compensate for non-uniform projector illumination across a flat build surface. In each of these cases, grayscale modulation is applied within the context of fixed-axis, planar layer printing, where layer thickness is uniform by definition.

The multi-axis DLP setting introduces a fundamentally different challenge. When the build platform is dynamically reoriented, sliced layers are non-planar and exhibit spatially varying thickness across their cross-section. Applying a uniform exposure time to such a layer will overcure thinner regions and undercure thicker ones, compromising dimensional accuracy and inter-layer adhesion. None of the prior grayscale works addresses this scenario. This paper bridges that gap by introducing a cure map generation method that combines spatially resolved grayscale modulation and exposure time control specifically for non-uniform layers produced by a multi-axis DLP slicing pipeline, enabling accurate curing across non-planar layer profiles without increasing layer count or requiring hardware modification. This approach is demonstrated on a set of representative 3D models with application toward the reproduction of automotive components.

\begin{figure}[H]
\centering
\includegraphics[width=\linewidth]{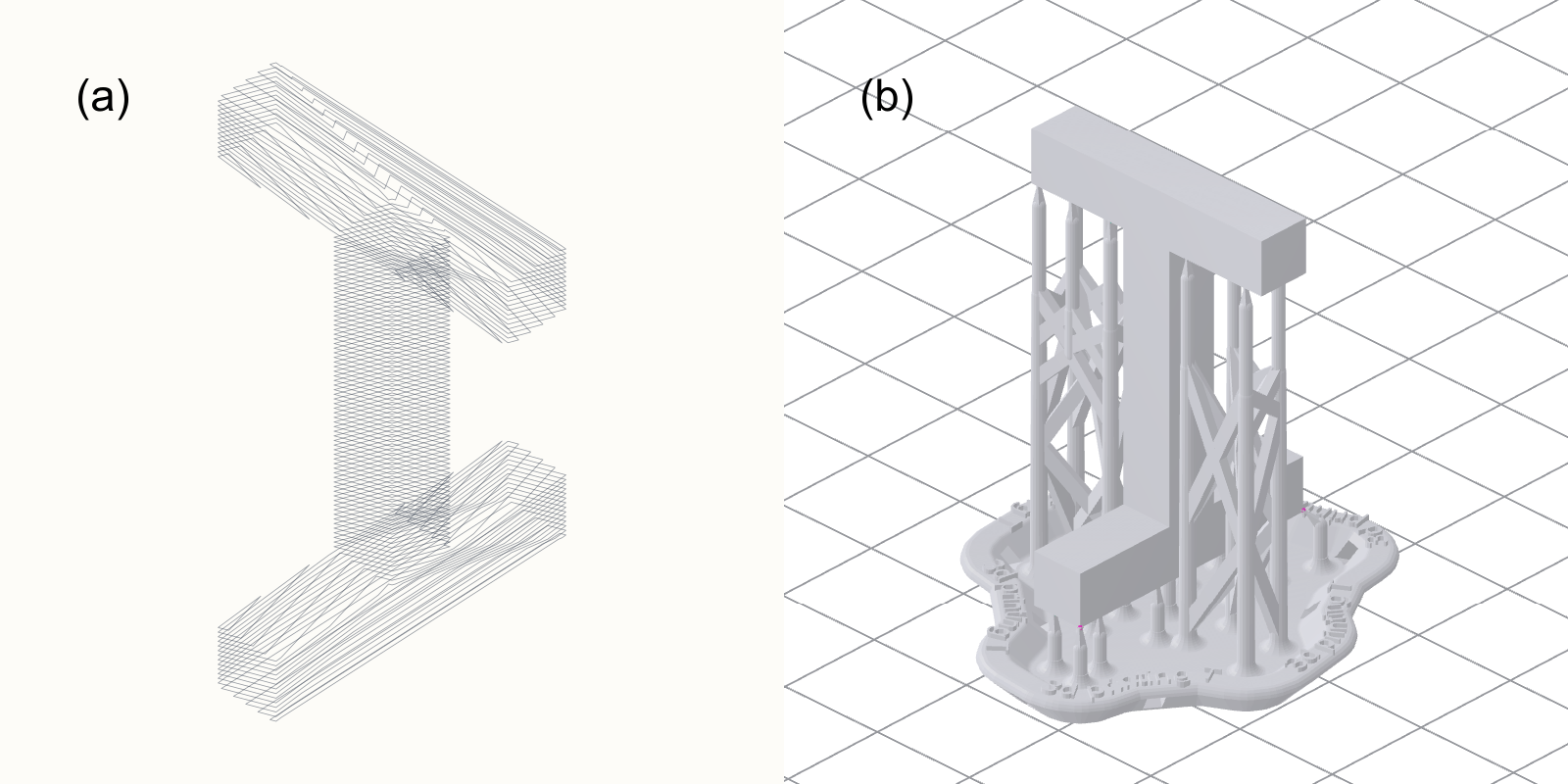}
\caption{Comparison of support structures for the T-shape object: (a) multi-axis curved slicing with minimal supports; (b) default slicing in Preform with extensive support structures required to sustain overhanging geometry.}\label{fig:support_comparison}
\end{figure}

\section{Methodology}
\subsection{Key Contribution}
The primary contribution of this work is the generation of spatially resolved cure maps for non-planar layers produced by a multi-axis DLP slicing pipeline. While prior work \cite{fang2020reinforced} established the broader framework for multi-axis DLP printing, the computation of per-pixel grayscale exposure masks, which encode the exact UV dose required at each point of a sliced layer to achieve a target cure depth, was not fully developed. This paper addresses that gap by introducing a two-stage algorithm: a curved iso-surface slicing algorithm that decomposes an input 3D mesh into a sequence of non-planar contour layers, followed by a cure-depth mask generation algorithm that converts each layer contour into a grayscale exposure map suitable for direct input to a DLP projector.

\subsection{Slicing Algorithm}
The slicing algorithm is adapted from the multi-axis FDM framework introduced by Fang et al. \cite{fang2020reinforced}, which decomposes solid models into collision-free working surfaces via iso-surfaces of an optimized scalar field. The original algorithm was converted to pseudocode and subsequently re-implemented in Python to accommodate the discrete, layer-based exposure requirements of DLP printing rather than the continuous toolpaths of FDM.

Given an input triangle mesh $M = (V, F)$ and a desired layer count $N$, the algorithm constructs a scalar field $\phi(x, y, z)$ over the mesh vertices. Two field modes are supported: a warped-$z$ field, which introduces sinusoidal perturbations to the vertical coordinate to produce curved layers, and a radial field, which generates concentric spherical layers about a user-defined center point. The scalar range $[\phi_{min}, \phi_{max}]$ is divided into $N$ evenly spaced intervals, and each layer $k$ is extracted as the iso-surface at value:

\begin{equation}
    \text{iso}_k = \phi_{min} + \left(k + 0.5\right) \cdot \frac{\phi_{max} - 
    \phi_{min}}{N}
\end{equation}

The midpoint placement of iso-values avoids sampling at scalar bounds and improves numerical stability. For each triangle in $F$, edge crossings at $\text{iso}_k$ are computed via linear interpolation and assembled into closed polyline contours $P_k$. Each layer is exported as an 3D contour file alongside a summary JSON file containing per-layer statistics.

A key constraint imposed on the slicing process is the enforcement of minimum and maximum layer thickness bounds. Layers falling outside these bounds are flagged or subdivided to ensure that the resulting cure maps remain within the operable range of the DLP projector and resin material system.

The pseudocode for the slicing algorithm is presented in Algorithm~\ref{alg:slice}.The algorithm relies on three subroutines. 
\textsc{BuildScalarField} constructs the scalar value $\phi$ at each mesh vertex according to the selected field mode, either applying sinusoidal perturbations to the $z$ coordinate for the warped-$z$ mode or computing Euclidean distance to a center point for the radial mode. 
\textsc{TriangleIsoSegments} processes each triangle by computing signed distances from its vertices to the current iso-value, detecting edge crossings via sign bracketing, and returning interpolated crossing points as a line segment. Crossing points that fall within the welding tolerance $\delta_{weld}$ of one another are deduplicated to avoid geometric artifacts.
\textsc{BuildPolylines} then chains the resulting segments across all triangles into ordered, closed polyline contours $P_k$ by matching shared endpoints within the same tolerance.

\begin{algorithm}
\caption{Curved Iso Slice OBJ}
\label{alg:slice}
\begin{algorithmic}[1]
\Require Triangle mesh $M = (V, F)$, layer count $N$, field parameters, 
tolerances $\varepsilon$, $\delta_{weld}$
\Ensure Polyline contour files and summary JSON
\State $\phi$ = \Call{BuildScalarField}{$V$, field\_params}
\State $lo  = \min(\phi)$; $hi \ = \max(\phi)$
\State $step = (hi - lo) / N$
\For{$k = 0$ to $N-1$}
    \State $iso_k = lo + (k + 0.5) \cdot step$
    \State $S_k = \emptyset$
    \For{each triangle $(a, b, c) \in F$}
        \State $segs$ = \Call{TriangleIsoSegments}{$(V_a, V_b, V_c)$, 
        $(\phi_a, \phi_b, \phi_c)$, $iso_k$, $\varepsilon$, $\delta_{weld}$}
        \State Append $segs$ to $S_k$
    \EndFor
    \State $P_k = $ \Call{BuildPolylines}{$S_k$, $\delta_{weld}$}
    \State \Call{WriteLayerOBJ}{$k$, $P_k$}
\EndFor
\State Write layers\_summary.json
\end{algorithmic}
\end{algorithm}

\subsection{Cure Map Generation}
Once each layer contour $P_k$ is obtained, a per-pixel grayscale exposure mask is computed using the Beer-Lambert cure depth model \cite{liu2026support}. The governing relationship between grayscale command $G(u,v)$, effective irradiance $I(u,v)$, and target cure depth $h(u,v)$ at projector pixel coordinates $(u,v)$ is:

\begin{equation}
    I(u,v) = I_0 \cdot \frac{G(u,v)}{255} \label{beer-lambert}
\end{equation}

\begin{equation}
    h(u,v) = D_p \ln\left(\frac{I_0 \cdot \frac{G(u,v)}{255} \cdot t_e}{H_e}\right)
\end{equation}

Inverting to solve for the required grayscale value given a target thickness profile $h(u,v)$:

\begin{equation}
    G(u,v) = \left\lfloor \frac{255 \cdot H_e}{I_0 \cdot t_e} 
    \exp\left(\frac{h(u,v)}{D_p}\right) \right\rfloor \label{greyscalevalue}
\end{equation}

The layer contour is first rasterized into a binary interior map using an even-odd rule. For each interior pixel, the target thickness $h(u,v)$ is evaluated, the required grayscale value is computed, clamped to $[0, 255]$ to satisfy projector bounds, and the result is written to a PGM image file. A saturation metric, the ratio of pixels at grayscale extremes to total exposed pixels, is tracked as a quality indicator; high saturation suggests miscalibration of material or process parameters.

The pseudocode for cure map generation is presented in Algorithm~\ref{alg:cure}. The algorithm relies on two subroutines. 
\textsc{ProjectToImageAndCloseLoop} maps each world-coordinate polyline contour $P_k$ into image pixel space using the provided world-to-image transform, closing the loop to ensure a valid boundary for rasterization. 
\textsc{RasterizeEvenOdd} fills the projected loops using an even-odd scanline rule, producing a binary interior map in which pixels inside the layer boundary are marked for exposure and all others are set to zero.

\begin{algorithm}
\caption{Generate CureDepth Mask}
\label{alg:cure}
\begin{algorithmic}[1]
\Require Layer contour $P_k$, raster size $W \times H$, projection transform, 
thickness profile $h(u,v)$, parameters $D_p, I_0, t_e, H_e$
\Ensure Grayscale mask image and statistics
\State $inside  = $ \Call{RasterizeEvenOdd}{$P_k$, $W$, $H$}
\For{$y  = 0$ to $H-1$}
    \For{$x  =  0$ to $W-1$}
        \If{$inside[y,x] = 0$}
            \State $img[y,x] =  0$; \textbf{continue}
        \EndIf
        \State $u  = (x+0.5)/W$; $v  = (y+0.5)/H$
        \State $h = $ thickness\_profile$(u, v)$
        \State $G_{raw} = \frac{255 \cdot H_e}{I_0 \cdot t_e} 
        \exp\!\left(\frac{h}{D_p}\right)$
        \State $img[y,x] = \text{clamp}(\text{round}(G_{raw}), 0, 255)$
    \EndFor
\EndFor
\State \Call{WritePGM}{$img$, $W$, $H$}
\end{algorithmic}
\end{algorithm}

\section{Light Exposure Time}

\subsection{Cure Depth and Photoexposure}
Accurate control of cure depth is fundamental to the quality and dimensional fidelity of DLP 3D printed parts. In vat photopolymerization, the depth to which resin is cured by a UV exposure is governed by the Beer-Lambert law \cite{swinehart1962beer}, which describes the attenuation of light intensity as it propagates through an absorbing medium. For photopolymer resins, this relationship is captured by the Jacobs working curve \cite{lee2001cure}: \begin{equation} C_d = D_p \ln\left(\frac{H_e}{H_0}\right)\end{equation}where $C_d$ is the cure depth, $D_p$ is the penetration depth of the resin, $H_e$ is the photoexposure dose delivered to the resin surface, and $H_0$ is the critical exposure dose required to initiate polymerization. The photoexposure dose is the product of light intensity and exposure duration:\begin{equation}H_e = I_0 \cdot t_e\end{equation}where $I_0$ is the peak irradiance of the projector and $t_e$ is the exposure time. Substituting, the cure depth can be expressed directly as a function of exposure time:

\begin{equation}
    C_d = D_p \ln\left(\frac{I_0 \cdot t_e}{H_0}\right)
\end{equation}

\subsection{Importance of Spatially Varying Exposure}
\begin{figure}[h]
\centering
\includegraphics[width=0.6\linewidth]{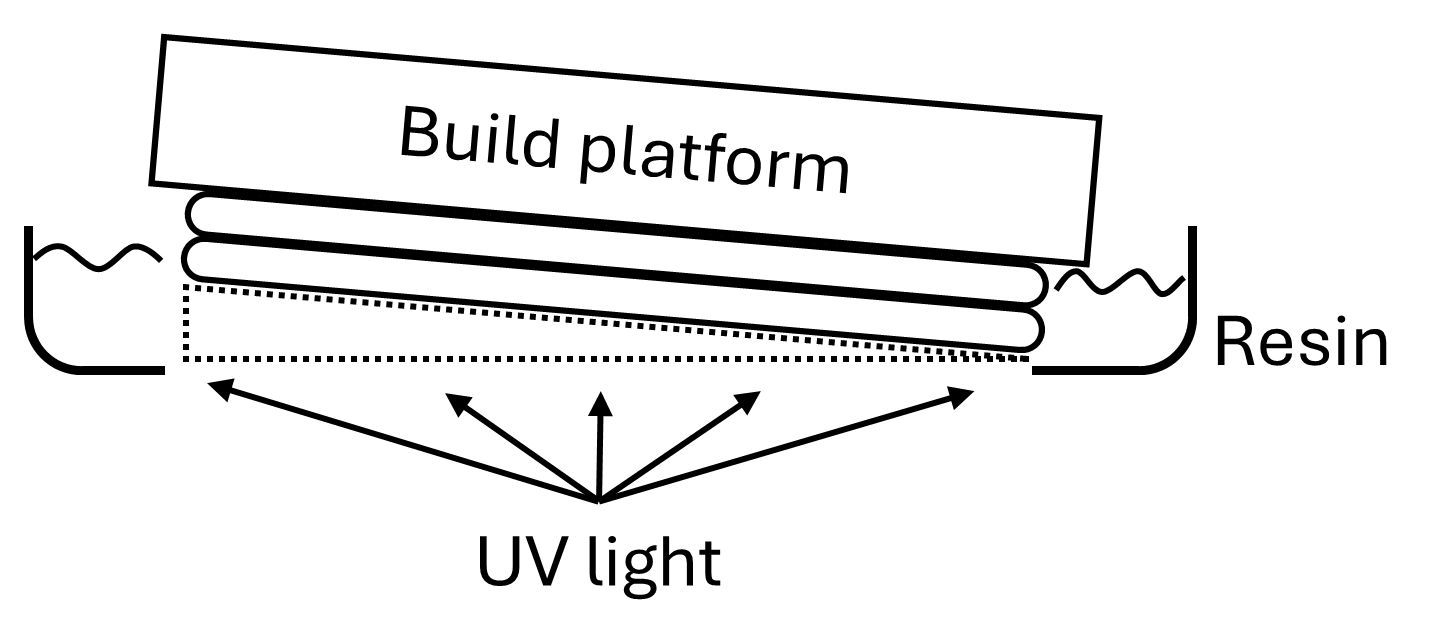}
\caption{Non-uniform layer thickness of multi-axis DLP printing.}\label{fig:depth}
\end{figure}

In conventional planar DLP printing, all pixels within a layer are exposed for the same duration at uniform intensity, which is appropriate when each layer has consistent thickness throughout. However, as shown in Figure \ref{fig:depth}, in multi-axis DLP printing, the dynamic reorientation of the build platform results in sliced layers with spatially non-uniform thickness \cite{liu2026support}. Applying a single uniform exposure time to such a layer will overcure thinner regions and undercure thicker ones, leading to dimensional inaccuracy, inter-layer adhesion failures, and degraded surface quality.

To address this, the exposure must be modulated spatially across each layer to match the local thickness requirement at every point. This is the motivation for generating a per-pixel cure map rather than a single scalar exposure time per layer. By encoding the required grayscale value at each projector pixel, the cure map ensures that each region of the layer receives precisely the dose needed to achieve the target cure depth, regardless of how thickness varies across the layer cross-section.

\subsection{Exposure Area and Time Calculation}
Given a target thickness profile $h(u,v)$ defined over the layer cross-section in normalized projector coordinates $(u,v)$, the required grayscale command at each pixel is derived by inverting the cure depth equation from earlier \eqref{greyscalevalue} 

The exposed area of layer $k$ is computed from the rasterized interior map. If $N_{exp,k}$ denotes the number of interior pixels and $p$ is the pixel pitch of the projector in millimeters per pixel, the physical exposed area is:

\begin{equation}
    A_k = N_{exp,k} \cdot p^2
\end{equation}

For a fixed exposure time $t_e$, the nominal energy delivered to the layer is 
\begin{equation} H_e = I_0 \cdot t_e\end{equation}
where process constraints permit, $t_e$ may be adapted per grayscale bin or region to extend the effective dynamic range of achievable cure depths beyond what a single exposure time would allow. The saturation ratio:\begin{equation}\sigma_k = \frac{N_{sat,k}}{N_{exp,k}}\end{equation}where $N_{sat,k}$ is the number of pixels at grayscale extremes ($G = 0$ or $G = 255$), serves as a calibration quality metric. A high saturation ratio indicates that the target thickness profile $h(u,v)$ is not achievable within the current process parameters $\{D_p, I_0, t_e, H_0\}$ and that recalibration is required \cite{maines2021sustainable}.

\section{Implementation}

\subsection{Pipeline Overview}
The implementation takes a triangle mesh in 3D format as input and produces two categories of output for each layer: a geometric contour encoding the three-dimensional position of the slice, and a grayscale cure mask encoding the required UV exposure at each projector pixel. The full pipeline is implemented in Python, adapted from the C++/Qt ReinforcedFDM framework introduced by Fang et al. \cite{fang2020reinforced}.

The end-to-end pipeline proceeds as follows. First, the input 3D file is parsed and any n-gon faces are fan-triangulated into triangles, producing an in-memory triangle mesh $M = (V, F)$. A scalar field $\phi(x, y, z)$ is then constructed over the mesh vertices according to the selected field mode. The scalar range $[\phi_{min}, \phi_{max}]$ is divided into $N$ evenly spaced iso-values, and each layer is extracted as the iso-surface contour at its corresponding iso-value. Finally, for each extracted contour, a grayscale cure mask is generated and written to disk alongside per-layer statistics.

\subsection{Position Encoding}
The three-dimensional position of each sliced layer is encoded geometrically and explicitly. For each layer $k$, edge crossings at iso-value $iso_k$ are detected by testing whether the scalar values at the two endpoints of a given edge bracket the iso-value:

\begin{equation}
    (iso_k - \phi_a)(\phi_k - \phi_b) < 0
\end{equation}

The crossing point is then computed by linear interpolation:

\begin{equation}
    \alpha = \frac{iso_k - \phi_a}{\phi_b - \phi_a}, \quad 
    \mathbf{p} = (1 - \alpha)\mathbf{p}_a + \alpha \mathbf{p}_b
\end{equation}

Crossing points across all triangles are welded by a tolerance $\delta_{weld}$ and chained into closed polyline contours. Each layer is exported as a separate 3D file containing the full set of three-dimensional contour vertices and ordered line connectivity records. The collection of all layer 3D files represents the sliced geometry as a stack of non-planar contours, suitable for robot arm path planning and exposure scheduling.

Two scalar field modes are supported. The warped-$z$ field introduces sinusoidal perturbations to the vertical coordinate to produce curved, non-planar layers:\begin{equation}\phi(x,y,z) = z + A\left(\sin(2\pi c_x \hat{x}) + \sin(2\pi c_y \hat{y})\right)\end{equation}where $\hat{x}$ and $\hat{y}$ are bounding-box normalized coordinates and $A$, $c_x$, $c_y$ are user-defined amplitude and cycle parameters. The radial field mode generates concentric spherical layers about a user-defined center point $\mathbf{c}$:

\begin{equation}
    \phi(x,y,z) = \lVert \mathbf{p} - \mathbf{c} \rVert_2
\end{equation}

\subsection{Light Exposure Encoding}
The UV exposure required at each point of a layer is encoded as a per-pixel grayscale value in a PGM image file, referred to as the cure mask. One cure mask is generated per layer and is sized to match the resolution of the DLP projector.

The layer contour is rasterized into a binary interior map as described in Algorithm~\ref{alg:cure}, with pixels outside the contour boundary assigned $G = 0$. 

For each interior pixel at normalized coordinates $(u, v)$, the target cure depth $h(u, v)$ is evaluated from the thickness profile, and the required grayscale command is computed by inverting the cure depth model as described in equation\eqref{greyscalevalue}.


A run-level summary file records per-layer diagnostics including the number of exposed pixels, grayscale minimum and maximum values, and the number of saturated pixels, providing a practical calibration quality check as described in Section~3.

\section{Results}
\subsection{Curved Slicing of 3D Models}
\begin{figure}[h]
\centering
\includegraphics[width=0.5\linewidth]{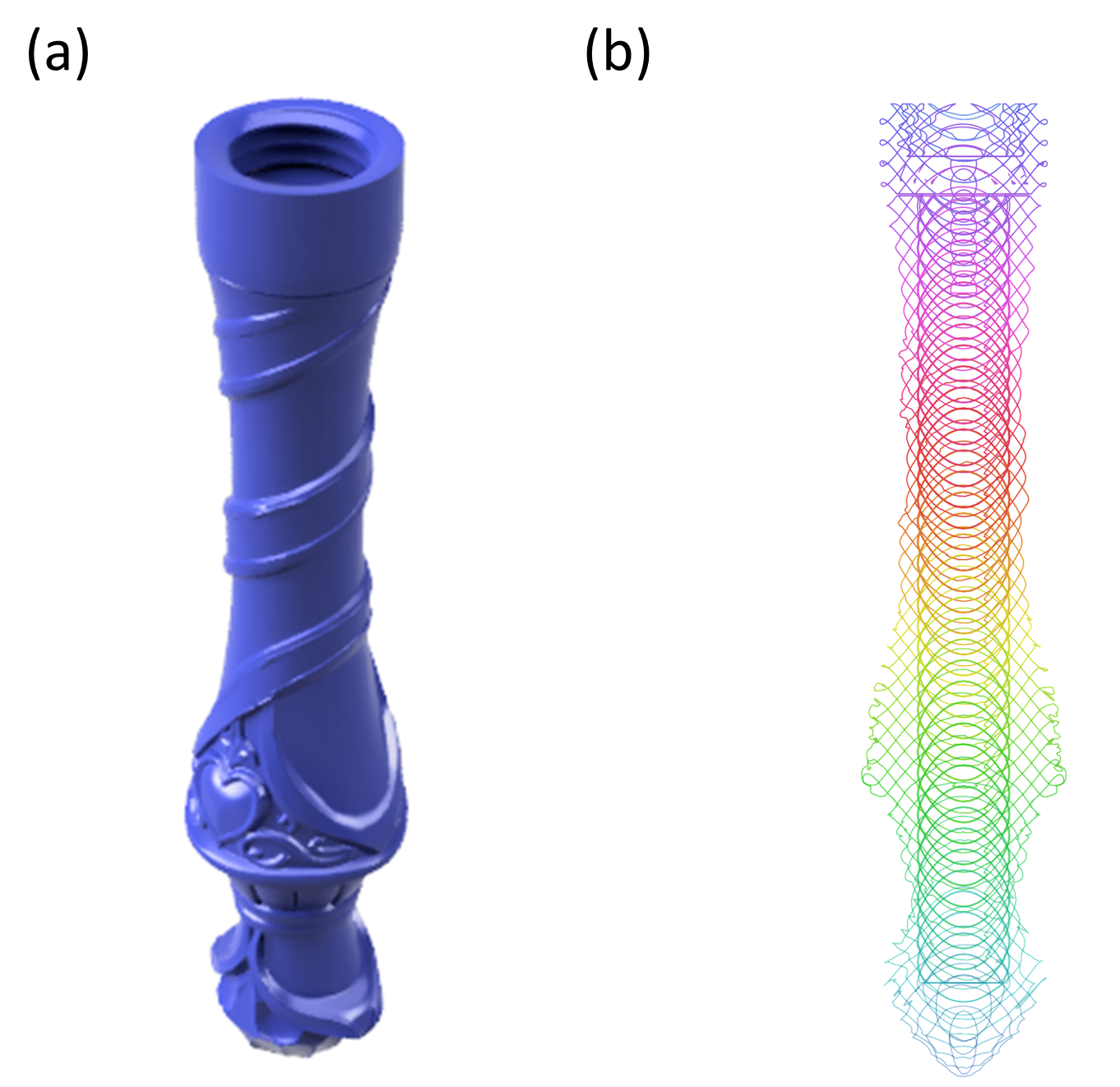}
\caption{Handle with delicate structure on the surface: (a) 3D file; (b) Curved sliced contour stack.}\label{fig:slice1}
\end{figure}

The slicing pipeline was applied to a set of representative 3D models to validate the curved iso-surface extraction algorithm. for Figure \ref{fig:slice1}(a) is the 3D model for the handle with delicate structure on surface, which is different for traditional DLP 3D printing to preserve all the features. Figure~\ref{fig:slice1}(b) shows the sliced contour stack, generated using the warped-$z$ scalar field with 200 layers. The non-planar nature of the extracted contours is visible in the curvature of individual slices across the object geometry. 


\begin{figure}[h]
\centering
\includegraphics[width=0.8\linewidth]{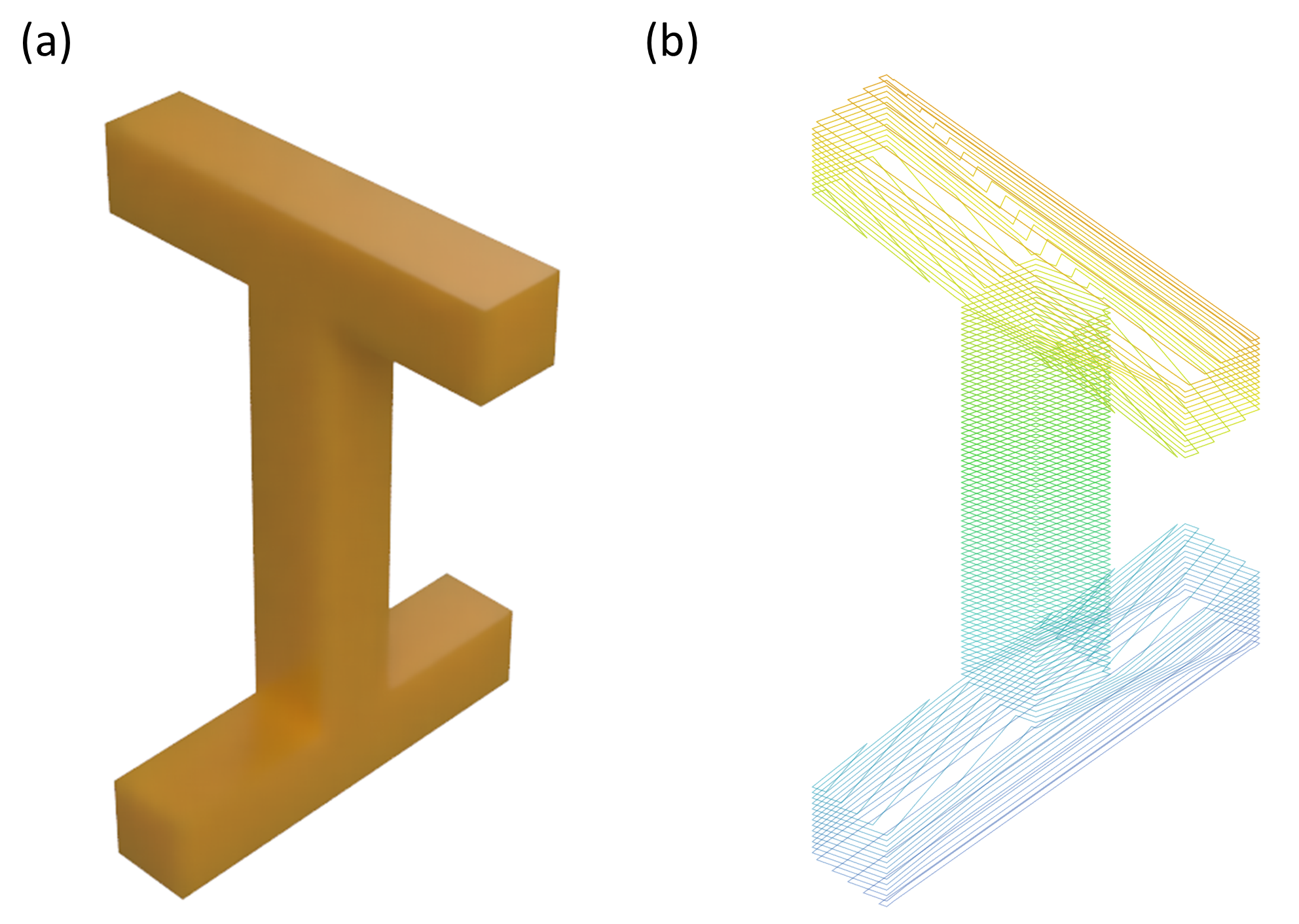}
\caption{T-shape object: (a) 3D file; (b) Curved sliced contour stack.}\label{fig:slice2}
\end{figure}



Figure~\ref{fig:slice2}(a) shows a T-shape object, which will be impossible to print without support structures with traditional DLP 3D printing. While Figure~\ref{fig:slice2}(b) shows the slicing of multi-axis slicing for the T-shape object, demonstrating that the algorithm generalizes across objects of varying geometric complexity.

\subsection{Cure Mask Generation}
\begin{figure}[h]
\centering
\includegraphics[width=\linewidth]{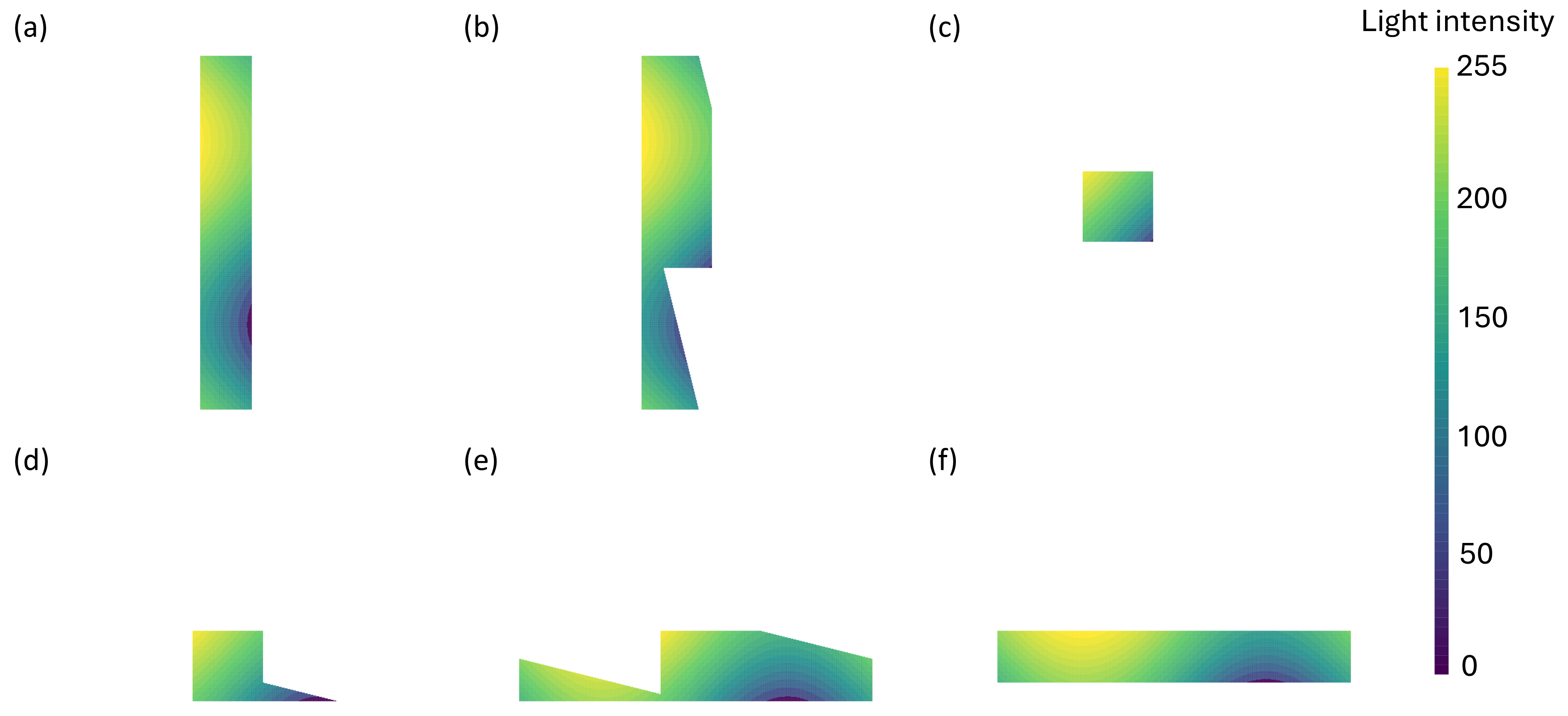}
\caption{Light intensity for selected layers for T-shape object: (a) layer 3; (b) layer 14; (c) layer 40; (d) layer 80; (e) layer 84; (e) layer 96.
}\label{fig:curemask}
\end{figure}

Figure~\ref{fig:curemask} shows representative cure masks generated for individual layers of Figure \ref{fig:slice2}. Each mask is a grayscale PGM image in which pixel intensity encodes the required UV exposure at that location, derived from the target thickness profile $h(u,v)$ using the inverted Beer-Lambert cure depth model. Brighter pixels correspond to regions requiring greater cure depth, while pixels outside the layer contour boundary are set to zero. The spatial variation in grayscale values across the mask reflects the non-uniform thickness of the corresponding sliced layer, confirming that the cure map correctly modulates exposure in response to local geometry.


 \subsection{Support Structure Reduction}
A key motivation for the multi-axis DLP approach is the elimination or reduction of support structures required during fabrication. To quantify this benefit, the sliced outputs were compared against conventional planar slicing baselines produced by Preform, a widely used SLA slicing tool that relies on fixed-axis, planar layer generation, using its default support generation settings as a consistent baseline. 

Table~\ref{tab:supports} summarizes the comparison for 2 test models. The reduction in layer count and resin volume observed in both models stems primarily from the multi-axis slicing and platform reorientation strategy, which aligns overhanging features with the gravity vector and eliminates the need for sacrificial support structures. The proposed cure map method plays a complementary but distinct role: it enables accurate fabrication of the non-uniform layers that result from this reorientation, ensuring that each region of a non-planar layer receives the correct UV dose to achieve the target cure depth. Without spatially resolved exposure control, the geometric benefits of multi-axis slicing could not be reliably realized in practice.

For Fig.~\ref{fig:slice1}, the proposed method achieved a 47\% reduction in resin usage compared to Preform. For Fig.~\ref{fig:slice2}, the proposed method achieved a 97\% reduction in support structure volume compared to Preform, consistent with the resin savings reported in the prior simulation study \cite{liu2026support}.

\begin{table}
\caption{Support structure comparison across slicing 
methods.}\label{tab:supports}
\centering{
\begin{tabular}{lp{3cm}p{2.5cm}} 
\toprule
Model & Proposed method & Preform \\
\midrule
Handle (Fig.~\ref{fig:slice1}) & 100 layers, 12.52~mL resin & 3176 layers, 20.63~mL resin\\
T-shape (Fig.~\ref{fig:slice2}) & 100 layers, 0.019~mL resin& 290 layers, 0.62~mL resin\\
\bottomrule
\end{tabular}
}
\end{table}
\subsection{Limitations}
The current implementation does not account for collision between the build platform and the resin vat boundary during multi-axis reorientation. While the slicing and cure map generation pipeline produces geometrically valid layers, ensuring that the robot arm trajectory avoids physical interference with the vat walls during fabrication is not yet integrated into the system. Addressing this through a collision-free path planning module is identified as a priority for future work, and would be a necessary step toward full end-to-end fabrication on a physical multi-axis DLP system.
\section{Conclusion}
This paper addressed the challenge of non-uniform layer thickness in multi-axis DLP 3D printing, where dynamic reorientation of the build platform produces sliced layers that vary in cure depth across their cross-section. To address this, a two-stage pipeline was developed: a curved iso-surface slicing algorithm adapted from the multi-axis FDM framework of Fang et al.~\cite{fang2020reinforced}, and a per-pixel cure map generation algorithm grounded in the Beer-Lambert cure depth model. Applied to a set of representative 3D models, the pipeline successfully produces non-planar contour layers and spatially resolved grayscale exposure masks, demonstrating that variable UV exposure can be encoded at pixel resolution without increasing total layer count or requiring hardware modification. This cure map approach represents a practical extension to existing multi-axis DLP 3D printing pipelines and has direct relevance to the fabrication of geometrically complex automotive components, where support structure reduction and geometric fidelity are critical manufacturing objectives. Future work will focus on experimental validation of the pipeline on a physical multi-axis DLP 3D printer, and integration with robot arm path planning to generate complete executable output for end-to-end fabrication.


\section*{Acknowledgments}
This research is partially supported from the University of Michigan Office of the Vice President for Research through Bold Challenges and AIIM Seed Networking Award.


\nocite{*}

\bibliographystyle{asmeconf}  
\bibliography{asmeconf-sample}

@article{wu2019mechanics,
  title={Mechanics of shape distortion of DLP 3D printed structures during UV post-curing},
  author={Wu, Dong and Zhao, Zeang and Zhang, Qiang and Qi, H Jerry and Fang, Daining},
  journal={Soft matter},
  volume={15},
  number={30},
  pages={6151--6159},
  year={2019},
  publisher={Royal Society of Chemistry}
}

@inproceedings{liu2022design,
  title={Design of Additively Manufactured Functionally Graded Cellular Structures},
  author={Liu, Zheng and Renteria, Anabel and Zheng, Zhuoyuan and Wang, Pingfeng and Li, Yumeng},
  booktitle={IIE Annual Conference. Proceedings},
  pages={1--6},
  year={2022}
}

@article{liu2025uncertainty,
  title={Uncertainty quantification of additively manufactured architected cellular materials for energy absorption applications},
  author={Liu, Zheng and Xu, Yanwen and Jiang, Yuan and Renteria, Anabel and Bansal, Parth and Xu, Chenlong and Wang, Pingfeng and Li, Yumeng},
  journal={ASCE-ASME Journal of Risk and Uncertainty in Engineering Systems, Part B: Mechanical Engineering},
  volume={11},
  number={3},
  pages={031204},
  year={2025},
  publisher={American Society of Mechanical Engineers}
}

@article{wang20203d,
  title={3D printing of viscoelastic suspensions via digital light synthesis for tough nanoparticle--elastomer composites},
  author={Wang, Kaiyang and Pan, Wenyang and Liu, Zheng and Wallin, Thomas J and van Dover, Geoffrey and Li, Shuo and Giannelis, Emmanuel P and Menguc, Yigit and Shepherd, Robert F},
  journal={Advanced Materials},
  volume={32},
  number={25},
  pages={2001646},
  year={2020}
}

@article{li2021digital,
  title={Digital light processing of liquid crystal elastomers for self-sensing artificial muscles},
  author={Li, Shuo and Bai, Hedan and Liu, Zheng and Zhang, Xinyue and Huang, Chuqi and Wiesner, Lennard W and Silberstein, Meredith and Shepherd, Robert F},
  journal={Science Advances},
  volume={7},
  number={30},
  pages={eabg3677},
  year={2021},
  publisher={American Association for the Advancement of Science}
}

@article{liu2022acoustophoretic,
  title={Acoustophoretic liquefaction for 3D printing ultrahigh-viscosity nanoparticle suspensions},
  author={Liu, Zheng and Pan, Wenyang and Wang, Kaiyang and Matia, Yoav and Xu, Artemis and Barreiros, Jose A and Darkes-Burkey, Cameron and Giannelis, Emmanuel P and Meng{\"u}{\c{c}}, Yi{\u{g}}it and Shepherd, Robert F and others},
  journal={Advanced Materials},
  volume={34},
  number={7},
  pages={2106183},
  year={2022}
}

@inproceedings{liu2026support,
  title={Support-Free Additive Manufacturing via Multi-Axis Digital Light Processing},
  author={Liu, Zheng},
  booktitle={AIAA SCITECH 2026 Forum},
  pages={0046},
  year={2026}
}

@article{fang2020reinforced,
  title={Reinforced FDM: Multi-axis filament alignment with controlled anisotropic strength},
  author={Fang, Guoxin and Zhang, Tianyu and Zhong, Sikai and Chen, Xiangjia and Zhong, Zichun and Wang, Charlie CL},
  journal={ACM Transactions on Graphics (TOG)},
  volume={39},
  number={6},
  pages={1--15},
  year={2020},
  publisher={ACM New York, NY, USA}
}

@article{patel2017highly,
  title={Highly stretchable and UV curable elastomers for digital light processing based 3D printing},
  author={Patel, Dinesh K and Sakhaei, Amir Hosein and Layani, Michael and Zhang, Biao and Ge, Qi and Magdassi, Shlomo},
  journal={Advanced Materials},
  volume={29},
  number={15},
  pages={1606000},
  year={2017}
}

@inproceedings{jumbo2021digital,
  title={Digital model to predict failures of porous structures in DLP-based additive manufacturing},
  author={Jumbo-Jaramillo, Ivannova and Lara-Padilla, Hernan},
  booktitle={XV Multidisciplinary International Congress on Science and Technology},
  pages={219--228},
  year={2021},
  organization={Springer}
}

@article{maines2021sustainable,
  title={Sustainable advances in SLA/DLP 3D printing materials and processes},
  author={Maines, Erin M and Porwal, Mayuri K and Ellison, Christopher J and Reineke, Theresa M},
  journal={Green Chemistry},
  volume={23},
  number={18},
  pages={6863--6897},
  year={2021},
  publisher={Royal Society of Chemistry}
}

@article{piedra20213d,
  title={3D printing parameters, supporting structures, slicing, and post-processing procedures of vat-polymerization additive manufacturing technologies: A narrative review},
  author={Piedra-Casc{\'o}n, Wenceslao and Krishnamurthy, Vinayak R and Att, Wael and Revilla-Le{\'o}n, Marta},
  journal={Journal of Dentistry},
  volume={109},
  pages={103630},
  year={2021},
  publisher={Elsevier}
}

@article{dilberoglu2017role,
  title={The role of additive manufacturing in the era of industry 4.0},
  author={Dilberoglu, Ugur M and Gharehpapagh, Bahar and Yaman, Ulas and Dolen, Melik},
  journal={Procedia manufacturing},
  volume={11},
  pages={545--554},
  year={2017},
  publisher={Elsevier}
}

@article{dai2018support,
  title={Support-free volume printing by multi-axis motion},
  author={Dai, Chengkai and Wang, Charlie CL and Wu, Chenming and Lefebvre, Sylvain and Fang, Guoxin and Liu, Yong-Jin},
  journal={ACM Transactions on Graphics (TOG)},
  volume={37},
  number={4},
  pages={1--14},
  year={2018},
  publisher={ACM New York, NY, USA}
}

@article{bi2023strength,
  title={Strength-enhanced volume decomposition for multi-directional additive manufacturing},
  author={Bi, Danjie and Duan, Molong and Lau, Tak Yu and Xie, Fubao and Tang, Kai},
  journal={Additive Manufacturing},
  volume={69},
  pages={103529},
  year={2023},
  publisher={Elsevier}
}

@article{li2021multi,
  title={Multi-axis support-free printing of freeform parts with lattice infill structures},
  author={Li, Yamin and Tang, Kai and He, Dong and Wang, Xiangyu},
  journal={Computer-Aided Design},
  volume={133},
  pages={102986},
  year={2021},
  publisher={Elsevier}
}

@article{feng2021curved,
  title={Curved-layered material extrusion modeling for thin-walled parts by a 5-axis machine},
  author={Feng, Xiaojing and Cui, Bin and Liu, Yaxiong and Li, Lianggang and Shi, Xiaojun and Zhang, Xiaodong},
  journal={Rapid Prototyping Journal},
  volume={27},
  number={7},
  pages={1378--1387},
  year={2021},
  publisher={Emerald Publishing Limited}
}

@article{lee2001cure,
  title={Cure depth in photopolymerization: Experiments and theory},
  author={Lee, Jim H and Prud'Homme, Robert K and Aksay, Ilhan A},
  journal={Journal of Materials Research},
  volume={16},
  number={12},
  pages={3536--3544},
  year={2001},
  publisher={Cambridge University Press}
}

@article{swinehart1962beer,
  title={The beer-lambert law},
  author={Swinehart, Donald F},
  journal={Journal of chemical education},
  volume={39},
  number={7},
  pages={333},
  year={1962},
  publisher={ACS Publications}
}

@inproceedings{luongo2020microstructure,
  title={Microstructure control in 3D printing with digital light processing},
  author={Luongo, Andrea and Falster, Viggo and Doest, Mads Brix and Ribo, Macarena Mendez and Eir{\'\i}ksson, Ey{\th}{\'o}r R{\'u}nar and Pedersen, David Bue and Frisvad, Jeppe Revall},
  booktitle={Computer Graphics Forum},
  volume={39},
  number={1},
  pages={347--359},
  year={2020},
  organization={Wiley Online Library}
}

@article{valentinvcivc2017low,
  title={Low cost printer for DLP stereolithography},
  author={Valentin{\v{c}}i{\v{c}}, Jo{\v{s}}ko and Pero{\v{s}}a, Matej and Jerman, Marko and Sabotin, Izidor and Lebar, Andrej},
  journal={Strojni{\v{s}}ki vestnik-Journal of Mechanical Engineering},
  volume={63},
  number={10},
  pages={559--566},
  year={2017}
}

@article{yu2023high,
  title={High-accuracy DLP 3D printing of closed microfluidic channels based on a mask option strategy},
  author={Yu, Zhengdong and Li, Xiangqin and Zuo, Tongxing and Wang, Qianglong and Wang, Huan and Liu, Zhenyu},
  journal={The International Journal of Advanced Manufacturing Technology},
  volume={127},
  number={7},
  pages={4001--4012},
  year={2023},
  publisher={Springer}
}

@inproceedings{mostafa2017tolerance,
  title={Tolerance control using subvoxel gray-scale DLP 3D printing},
  author={Mostafa, Khaled and Qureshi, AJ and Montemagno, Carlo},
  booktitle={ASME International Mechanical Engineering Congress and Exposition},
  volume={58356},
  pages={V002T02A035},
  year={2017},
  organization={American Society of Mechanical Engineers}
}

\appendix




\end{document}